\documentclass[prl,amsmath,aps,superscriptaddress, twocolumn]{revtex4}
\usepackage{graphicx,xspace}
\usepackage{float}
\usepackage{amsmath}
\usepackage{amssymb}
\usepackage[normalem]{ulem} 
\usepackage{epsfig}
\usepackage{graphicx,psfrag,xspace}

\begin{document}

\title{Asymptotic behavior of self-affine processes in semi-infinite domains}
\author{Andrea Zoia}
\email{andrea.zoia@cea.fr}
\affiliation{CEA/Saclay, DEN/DM2S/SFME/LSET, B\^at.~454, 91191 Gif-sur-Yvette Cedex, France}
\author{Alberto Rosso}
\affiliation{CNRS - Universit\'e Paris-Sud, LPTMS, UMR8626 - B\^at.~100, 91405 Orsay Cedex, France}
\author{Satya N.~Majumdar}
\affiliation{CNRS - Universit\'e Paris-Sud, LPTMS, UMR8626 - B\^at.~100, 91405 Orsay Cedex, France}

\begin{abstract}
We propose to model the stochastic dynamics of a polymer passing through a pore (translocation) by means of a fractional Brownian motion, and study its behavior in presence of an absorbing boundary. Based on scaling arguments and numerical simulations, we present a conjecture that provides a link between the persistence exponent $\theta$ and the Hurst exponent $H$ of the process, thus sheding light on the spatial and temporal features of translocation. Furthermore, we show that this conjecture applies more generally to a broad class of self affine processes undergoing anomalous diffusion in bounded domains, and we discuss some significant examples.
\end{abstract}
\maketitle

\begin{figure}[t]
\centerline{\epsfxsize=7.0cm
\epsfbox{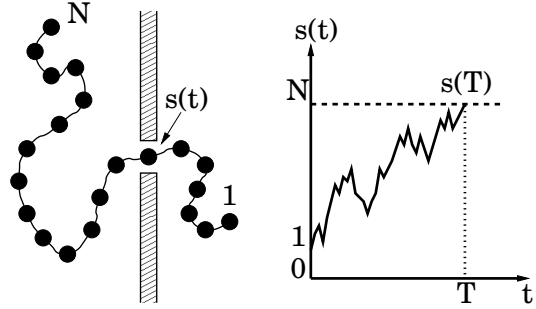}}
\caption{Left: Translocation of a polymer chain through a pore. Right: The translocation coordinate $s(t)$ denotes the number of the monomer that is crossing the pore at time $t$.}
   \label{fig0}
\end{figure}

The dynamics of a polymer chain composed of $N$ monomers passing through a pore (translocation) has been intensively investigated in recent years, by virtue of its central role in understanding, e.g., viral injection of DNA into a host or RNA transport through nanopores, and mastering such applications as fast DNA or RNA sequencing through engineered channels~\cite{ref0,ref3,ref5,ref1}. The translocation coordinate $s(t)$, namely the label of the monomer crossing the pore at time $t$, is key to quantitatively describing the translocation process~\cite{ref7,ref8,joanny}, which begins when $s=1$ and ends when $s=N$, i.e., when the first and the last monomer of the chain enter the pore, respectively (see Fig.~\ref{fig0}).

Various dynamical regimes of $s(t)$ have been identified: in absence of driving forces and hydrodynamic effects (free polymer), fluctuations dominate and $s(t)$ can be regarded as a stochastic process, whose features vary with the polymer length $N$~\cite{ref8,kardar3, kardar4}. Understanding the dynamics of $s(t)$ represents a challenging problem. A free polymer is characterized by two natural time scales. First, the intrinsic equilibration time $t_{eq}$ required by the center of mass of the polymer $R_{cm}$ to travel a distance of the order of the typical size of the chain. This size is given by the radius of gyration $R_{g}$, which scales as $R_g \sim N^\nu$ in the large $N$ limit. In a good solvent, $\nu=3/4$ in $2d$ and $\nu \simeq 0.59$ in $3d$ when exluded-volume effects for the monomers are considered; $\nu=1/2$ for an ideal (`phantom') polymer~\cite{degennes}. The center of mass diffuses with a diffusion coefficient $\sim 1/N$. Then, $R_{cm} \sim \sqrt{t/N}$. Hence, $\sqrt{t_{eq}/N} \sim R_g$ and $t_{eq} \sim N^{2\nu +1}$ for large $N$. On the other hand, the translocation time $T$ (much longer than $t_{eq}$) is the time required by the polymer to go through the pore, so that $s(T)=N$ (Fig.~\ref{fig0}). Under the hypothesis that the translocation is a self-affine process, i.e., $s(t) \sim t^H$, with Hurst exponent $H$, it follows that $T \sim N^{1/H}$.

For short polymers, excluded-volume effects are negligible, $s(t)$ undergoes diffusion, and $T \sim N^2$~\cite{ref8}. However, as $N$ increases, the excluded-volume interactions become relevant and $s(t)$ undergoes subdiffusion, $0<H<1/2$~\cite{kardar3}. Numerical simulations (mostly $2d$) support the following conclusions: $i)$ $T$ and $t_{eq}$ have the same scaling, up to a large prefactor, i.e., $T \sim t_{eq}$. Hence, $H=1/(1+2\nu)$~\cite{kardar3}; (Note that for `phantom' polymers $s(t)$ always diffuses, even for large $N$.) $ii)$ the probability $P(s,t)$ of finding the monomer $s$ in the pore at time $t$ for an infinite chain (i.e., in absence of boundaries) is Gaussian~\cite{kardar2}; $iii)$ for a finite chain yet to have completed translocation, $s(t)$ evolves in presence of two absorbing boundaries at $s=0$ and $s=N$ (see Fig.~\ref{fig0}) and the distribution of $s(t)$ converges to a non-Gaussian form at long times. In particular, this distribution vanishes nonlinearly as $s^\phi$ and $(N-s)^{\phi}$ at the two boundaries $s=0$ and $s=N$ respectively, with $\phi \simeq 1.44$ in $2d$~\cite{kardar2}. Computing analytically the exponent $\phi$is a nontrivial problem and is the main objective of this Letter.

Upon gathering these hints from simulations, we propose as a natural candidate for $s(t)$ the fractional Brownian motion (fBm), a self-affine Gaussian process ($0<H<1$). A Gaussian process is completely defined by its autocorrelation function $\langle s(t_1)s(t_2)\rangle$: for fBm, \cite{mandelbrot}
\begin{equation}
\langle s(t_1)s(t_2)\rangle \propto t_1^{2H}+ t_2^{2H}-|t_1-t_2|^{2H}.
\label{corr_r}
\end{equation} 
Brackets refer to ensemble average over many realizations. Equation~\ref{corr_r} implies that the incremental correlation function of fBm is stationary
\begin{equation}
\langle \left( s(t_1)- s(t_2) \right)^2 \rangle \propto |t_1-t_2|^{2H}.
\end{equation} 
Note that Brownian motion (BM) is fBm with $H=1/2$. When $H=1/(1+2\nu)$, fBm satisfies conditions $i)$ and $ii)$. Concerning $iii)$, the central result of our paper is to show that the probability distribution of a fBm confined in the positive half-axis vanishes as $s^\phi$, with
\begin{equation} 
\phi =\frac{1-H}{H},
\label{exponent}
\end{equation}
close to the absorbing boundary $s=0$, in the long time limit. For $2d$ polymers with excluded-volume effects, using $\nu=3/4$ one gets $H=2/5$ and $\phi=3/2$, which is in good agreement with the numerical value $\phi \simeq 1.44 $~\cite{kardar2}. In $3d$, using $\nu \simeq 0.59$, we predict $H \simeq 0.46$ and $\phi \simeq 1.18$.

More generally, we will show that actually for a broad class of self-affine processes with Hurst exponent $H$ the exponent $\phi$ is related to the persistence exponent $\theta$ of the process via the relation $\phi = \theta / H$. The persistence of a stochastic process is simply the probability of no return to its initial value up to time $t$ and for a wide class of processes the persistence decays algebraically $\sim t^{-\theta}$ at late times witha nontrivial persistence exponent $\theta$~\cite{maj2}. For fBm, in particular, $\theta =1-H$ is known exactly~\cite{majumdar}, leading to the result $\phi=(1-H)/H$. This general relation of $\phi$ to persistence exponent $\theta$ sheds light on the spatial and temporal features of anomalous diffusion in presence of absorbing boundaries, which is ubiquitous in nature and arises is such diverse fields as contaminant migration in heterogeneous materials~\cite{anomalous1} and charge transport in amorphous semiconductors, to name only a few~\cite{anomalous2}.

Let us begin by introducing the class of stochastic processes we are interested in. A generic unbiased stochastic walk $x(t)$ can be represented as a sum of jumps (increments) $\xi(t')$,
\begin{equation} 
x(t)=\sum_{t'=1}^{t} \xi(t') 
\end{equation} 
where the jump length $\xi(t')$ has a symmetric marginal distribution $\pi_{t'}(\xi)$. The Central Limit Theorem guarantees that, if $x(t)$ undergoes anomalous diffusion, one of the following statements must be violated: $a)$ the jumps are identically distributed, i.e., $\pi_{t'}(\xi)$ does not depend on time: $x(t)$ has stationary increments; $b)$ the variables $\xi$ are independent: $x(t)$ is Markovian; $c)$ $\pi_{t'}(\xi)$ has finite variance. For instance, fBm, which is the only self-affine Gaussian process with stationary increments, is anomalous and violates $b)$ (except for $H=1/2$). In the following, we present a general scaling argument for self-affine processes satisfying property $a)$, i.e., having stationary increments, but not restricted to satisfy either $b)$ or $c)$.

For such processes started in $x_0>0$ and killed upon leaving the positive half-axis, we define $P_+(x,x_0,t)$ as the probability of finding the walker in $x$ at time $t$. We define $p_{x_0}(x,t)$ as the conditional probability density of finding the walker, given that it has not been absorbed at any previous time:
\begin{equation}
p_{x_0}(x,t)=\frac{P_+(x,x_0,t)}{\int_{0}^{\infty} d x P_+(x,x_0,t)}.
\label{conditional}
\end{equation}
At long times, the small-$x$ behavior of this distribution gives access to the exponent $\phi$. The quantity $S(x_0, t)=\int_{0}^{\infty} d x P(x,x_0,t)$ defines the survival probability that the walker has not left the positive half-axis up to $t$.

As a useful guide, let us first recall the results for regular BM, i.e., fBm with $H=1/2$. The method of images gives the scaling form \cite{maj1}
\begin{equation}
P_+(x,x_0,t) = \frac{1}{\sqrt{t}} F \left(\frac{x}{\sqrt{t}}, \frac{x_0}{\sqrt{t}}\right),
\end{equation}
where $F(y,y_0)= [e^{-(y-y_0)^2/2}- e^{-(y+y_0)^2/2} ] / \sqrt{2 \pi}$. The long time behavior of the conditional probability, for any $x_0$, is given by $x_0 \to 0$:
\begin{equation}
p_{0}(x,t)= \frac{x}{\sqrt{t}}e^{-x^2/2t}.
\end{equation}
Observe that in this limit the process is not reminiscent of the initial condition. For large $x$, the particles do not feel the presence of the boundary, and $p_{0}(x,t)$ behaves as the Gaussian probability of an unconstrained BM. On the other hand, $p_{0}(x,t)$ vanishes linearly close to the boundary, which implies that $\phi = 1$ for BM.

Inspired by the special case of BM, we anticipate a similar scaling for a generic self-affine process:
\begin{equation}
P_+(x,x_0,t)= \frac{1}{t^{H'}} F \left(\frac{x}{t^{H}}, \frac{x_0}{t^{H}}\right) ,
\label{scaling_1}
\end{equation}
where $H'$ has to be determined. Integrating over $x$, we get $S(x_0,t)=t^{H-H'}G(x_0/t^{H})$, where $G(y) = \int_0^{\infty} F(x,y) dx$ is a scaling function of a single variable. For any $x_0>0$, we know that $S(x_0,0)=1$. Therefore, the only acceptable choice is $H'=H$, which implies
\begin{equation}
S(x_0, t)=G\left( \frac{x_0}{t^{H}}\right) ,
\label{s_g}
\end{equation}
with $G(+\infty)=1$. This scaling form allows the asymptotic behavior of $S(x_0,t)$ to be fully characterized. In particular, since the survival probability is expected to decay as $S(x_0,t)\sim t^{-\theta}$ for large $t$ (the semi-infinite domain lacks a characteristic scale), $G(y)$ must behave as $G(y)\sim y^{\theta/H}$ for small $y$, so that
\begin{equation}
S(x_0,t) \propto t^{-\theta}x_0^{\theta/H}.
\label{asy}
\end{equation}
Consider now the scaled variables $y=x/t^H$ and $y_0=x_0/t^H$. By combining Eq.~\ref{s_g} with Eqs.~\ref{scaling_1} and~\ref{conditional}, we get
\begin{equation}
F(y,y_0)=S(y_0)p_{y_0}(y)
\label{FY}
\end{equation}
where $p_{y_0}(y)$ is the conditional probability density expressed in terms of the rescaled variables. In the long time limit, $y_0 \to 0$ and $F(y,y_0)$ can be factorized as 
\begin{equation}
F(y,y_0) \propto y_0^{\theta/H} p_{0}(y).
\label{FY2}
\end{equation}
Let us now consider the limit $y \to 0$ and suppose that $p_{0}(y) \sim y^\phi$. The process is time reversible since its increments are stationary, i.e., a path from $x_0$ to $x$ forward in time plays the same role of a path from $x$ to $x_0$ backward in time. As a consequence, the two limits $x_0 \to 0$ and $x \to 0$ can be interchanged in $P_+(x,x_0,t)$, implying that $y_0$ and $y$ must appear in an identical fashion in the limiting behavior of $F(y,y_0)$ in Eq.~(\ref{FY2}), and thus follows our proposed scaling relation $\phi=\theta/H$. This conjecture represents our central result.
\begin{figure}[t]
\centerline{\epsfxsize=9.0cm\epsfbox{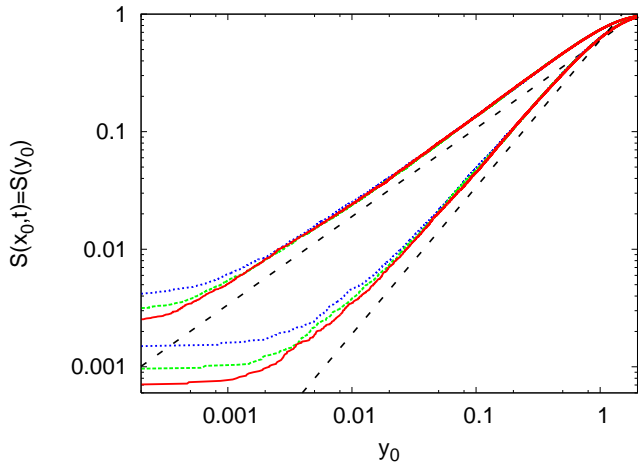}}
\caption{Survival probability as a function of the rescaled variable $y_0$, 
for $H=4/7$ (top) and $H=4/9$ (bottom), at various times $t$ in a semi-infinite domain. 
Numerical simulations are compared with the proposed scaling $(1-H)/H=3/4$ and $(1-H)/H=5/4$ 
respectively.}
\label{fig1}
\end{figure}

In support of this conjecture, we proceed next to numerically confirm its validity for fBm. In order to generate a fBm path of $L+1$ steps, $\left\lbrace x_0, x_1, ...,x_i, ... , x_{L} \right\rbrace $, we need to draw $L$ Gaussian numbers correlated through Eq.~\ref{corr_r}. This is in general a time consuming procedure, as it requires computing the square root of the autocorrelation matrix. Better results can be obtained by making use of the stationarity of the increments $\xi_i=x_i-x_{i-1}$, with a Toeplitz matrix correlator $\langle \xi_i \xi_{i+k} \rangle \propto |k-1|^{2H}+ |k+1|^{2H}-2|k|^{2H}$. In particular, the Levinson algorithm allows first passage problems to be efficiently tackled, by recursively generating $x_{i+1}$ given $x_i, \ldots, x_0$~\cite{hosking}.

In Fig.~\ref{fig1}, we test the scaling of $S(x_0,t)$ proposed in Eq.~\ref{asy}. The survival probability is shown as a function of the rescaled variable $y_0$, for different times. Particles start at $x_0>0$ and a single absorbing boundary is set in $x=0$. All curves collapse, in agreement with the self-affinity of the process. The collapse does not work in the region where $x_0$ is of the same order of magnitude as the typical increment of the (discrete) process. This effect disappears as time (and thus path length) increases, approaching the continuum limit. A good agreement is found between the proposed conjecture and the numerical results. In the superdiffusive regime, $t^H$ grows faster, and shorter times are required to observe the predicted scaling, i.e., $\phi=(1-H)/H$.

\begin{figure}[t]
\centerline{\epsfxsize=9.0cm
\epsfbox{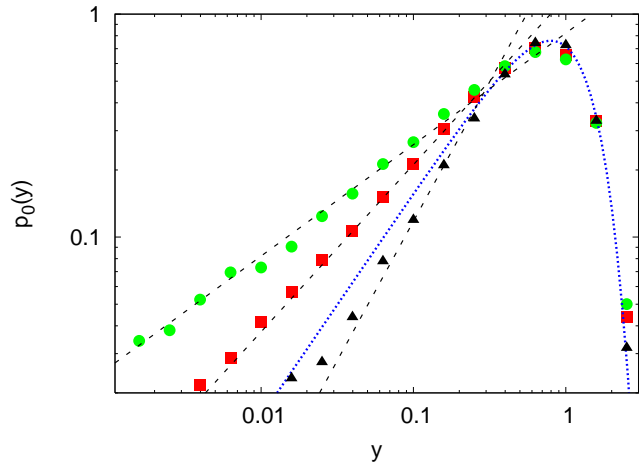}}
\caption{Semi-infinite domain. Conditional probability $p_0(y)$ as a function of the rescaled variable $y=x/ct^H$ for different Hurst exponents $H$. The constant $c$ is chosen so that $y$ has unit mean. From top to bottom, symbols are $H=2/3$ ($L=4 \, 10^3$), $H=4/7$ ($L=1.2 \, 10^4$) and $H=4/9$ ($L=2 \, 10^4$). Dashed lines correspond to the predicted slopes of Eq.~\ref{exponent}. Dotted line corresponds to $H=1/2$, where $p_0(y)=2y/\pi \exp(-\pi y^2/4)$.}
   \label{fig2}
\end{figure}

We proceed then to verify that the same scaling exponent $\phi$ characterizes also the behavior of the conditional probability close to the absorbing boundary. This is done in Fig.~\ref{fig2}, where $p_{0}(y)$ is plotted as a function of $y$ for some exponents $H$. For all curves, the rescaled variable $y_0$ varies in the range $0.1-0.4$, which a posteriori is verified to satisfy the long time limit. Good agreement with the proposed conjecture is found. Again, convergence to the expected scaling is faster in the superdiffusive case. Far from the boundary, $p_{0}(y)$ has a Gaussian decay, i.e., behaves as the free propagator, as observed above.

Now, we briefly review some stochastic transport processes where the proposed conjecture is shown to work.

{\em L\'evy Flights} are Markovian superdiffusive processes whose jumps obey a L\'evy stable (symmetric) law of index $0<\mu \le 2$~\cite{shlesinger,zoia_rosso}. Since $\pi(\xi) \sim \xi^{-1-\mu}$, property $c)$ is violated. The Hurst exponent is $H=1/\mu$. By virtue of the Sparre Andersen theorem~\cite{sparre}, the persistence exponent is $\theta=1/2$, independent of $\mu$. It has been shown that far from the origin $p_{0}(y)$ behaves as L\'evy stable law, whereas the conditional distribution close to the origin scales as $p_{0}(y) \sim y^{1/2H}$~\cite{zumofen}, hence the exponent $\phi=1/2H=\theta/H$.

Within the framework of the so-called {\em Continuous Time Random Walk} model, a class of subdiffusive walks is introduced by assuming that the waiting times between consecutive (Gaussian and symmetrical) jumps have a power-law decay of the kind $t^{-1-\alpha}$, with $0<\alpha<1$, so that the Hurst exponent is $H=\alpha/2$~\cite{klafter1}. This process is stationary, provided that the walker has made a jump exactly at initial time~\cite{lubelski}. In presence of an absorbing boundary, it has been shown that the survival probability behaves as $S(x_0,t) \sim x_0/t^{\alpha/2}$ for large times~\cite{absorbing_sub}, hence the persistence exponent reads $\theta=H$. Moreover, $p_{0}(y)$ vanishes linearly close to the boundary~\cite{absorbing_sub}, so that $\phi=1=\theta/H$.
\begin{figure}[t]
\centerline{\epsfxsize=9.0cm
\epsfbox{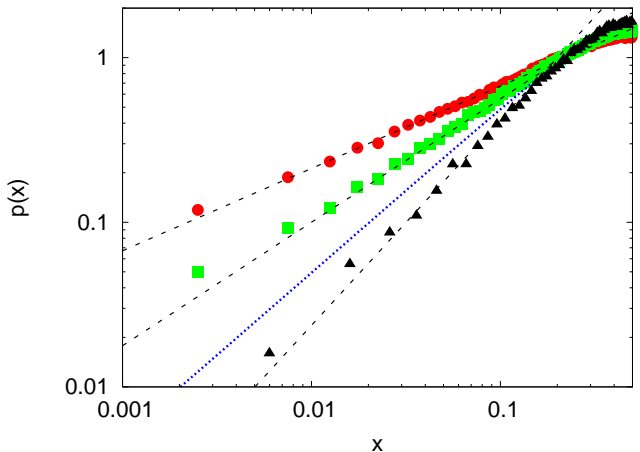}}
\caption{Bounded box of unit size. Conditional probability $p(x)$ as a function of $x$ for different Hurst exponents $H$. From top to bottom, symbols are $H=2/3$ ($L=4 \, 10^3$), $H=4/7$ ($L=1.2 \, 10^4$) and $H=4/9$ ($L=2 \, 10^4$). Dashed lines correspond to the predicted slopes of Eq. \ref{exponent}. Dotted line corresponds to $H=1/2$, where $p(x)=\pi\sin(\pi x)/2$.}
   \label{fig3}
\end{figure}

So far, we have focused on a semi-infinite region; several processes (including translocation) involve however finite-size domains~\cite{kardar3,kardar1}. Asymptotically, for large $N$ and close to the origin, the conditional probability in a bounded box must be independent of $t$ and $x_0$, i.e., $p_{x_0}(x,t)=p(x)$ and one would expect $p(x) \sim x^\phi$, where $\phi$ is the same exponent as the semi-infinite case: the particle feels the presence of the other boundary at $N$ only when sufficiently close to it. The case of BM $(H=1/2)$ in a bounded box can be analytically solved by addressing the associated Laplacian eigenvalue problem; it turns out that for long times $\phi=1$: the conditional probability close to the boundaries of the box has the same scaling as in a semi-infinite domain. A numerical test for fBm with other values of $H$ is provided in Fig.~\ref{fig3}, which again reveals a good agreement between simulation results and the proposed conjecture. The same holds true also for the other processes considered above.

Finally, one can also interpret our results via a simple heuristic argument. Subdiffusive processes tend to explore space more thoroughly than regular diffusion, so that close to the boundary their probability of being absorbed is larger. We therefore expect $\phi \ge 1$ for subdiffusion. As an example, excluded-volume effects make the translocation coordinate subdiffusive, and the untranslocated polymers are more easily found far from the boundary at any given time. Analogous arguments lead to the conclusion that $\phi \le 1$ for superdiffusion.

The authors thank P. Borgnat, M. Kardar and S. Roux for useful discussions.

\end{document}